\pdfoutput=1

\documentclass[10pt,prd,aps,amssymb,amsmath,tightenlines,showpacs,twocolumn]{revtex4}
\usepackage{graphicx}

\newcommand{\cP}{\ensuremath{\mathcal{P}}}
\newcommand{\cT}{\ensuremath{\mathcal{T}}}
\newcommand{\cPT}{\ensuremath{\mathcal{PT}}}

\makeatletter
\def\Ddots{\mathinner{\mkern1mu\raise\p@
\vbox{\kern7\p@\hbox{.}}\mkern2mu
\raise4\p@\hbox{.}\mkern2mu\raise7\p@\hbox{.}\mkern1mu}}
\makeatother

\begin{document}

\title{Twofold Transition in $\cPT$-Symmetric Coupled Oscillators}

\author{Carl M. Bender$^a$}\email{cmb@wustl.edu}
\author{Mariagiovanna Gianfreda$^b$}\email{Maria.Gianfreda@le.infn.it}
\affiliation{$^a$Department of Physics, Washington University, St. Louis, MO
63130, USA and Department of Mathematical Science, City University London,
Northampton Square, London EC1V 0HB, UK \\
$^b$Dipartimento di Matematica e Fisica Ennio De Giorgi,
Universit\`a del Salento and I.N.F.N. Sezione di Lecce, Via Arnesano, I-73100
Lecce, Italy}

\date{\today}

\begin{abstract}
The inspiration for this theoretical paper comes from recent experiments on a
$\cPT$-symmetric system of two coupled optical whispering galleries (optical
resonators). The optical system can be modeled as a pair of coupled linear
oscillators, one with gain and the other with loss. If the coupled oscillators
have a balanced loss and gain, the system is described by a Hamiltonian and the
energy is conserved. This theoretical model exhibits {\it two} $\cPT$
transitions depending on the size of the coupling parameter $\epsilon$. For
small $\epsilon$ the $\cPT$ symmetry is broken and the system is not in
equilibrium, but when $\epsilon$ becomes sufficiently large, the system
undergoes a transition to an equilibrium phase in which the $\cPT$ symmetry is
unbroken. For very large $\epsilon$ the system undergoes a second transition and
is no longer in equilibrium. The classical and the quantized versions of the
system exhibit transitions at exactly the same values of $\epsilon$.
\end{abstract}


\maketitle

\section{Introduction}
\label{s1}

The predicted properties of $\cPT$-symmetric Hamiltonians \cite{A1,A2} have been
observed at the classical level in a wide variety of laboratory experiments
involving superconductivity \cite{R1,R2}, optics \cite{R3,R4,R5,R6}, microwave
cavities \cite{R7}, atomic diffusion \cite{R8}, nuclear magnetic resonance
\cite{R9}, and coupled electronic and mechanical oscillators \cite{R10,R11}.
Although $\cPT$-symmetric systems were originally explored at a highly
mathematical level, it is now understood that one can interpret $\cPT$-symmetric
systems simply as nonisolated physical systems having a balanced loss and gain.

In this paper we examine a mathematical model based on recent unpublished
experiments. These experiments were performed on a system consisting of two
coupled $\cPT$-symmetric whispering galleries (optical resonators) \cite{R12}.
Such a system is $\cPT$ symmetric if one resonator is optically driven and the
other resonator has a balanced loss. We examine here the properties of the
mathematical model at a theoretical level and we study both the classical and
the quantum versions of the system.

A system of two identical coupled resonators, one with loss and the other with
gain, can be modeled as coupled oscillators whose amplitudes are $x(t)$ and
$y(t)$. Both oscillators have a natural frequency $\omega$. The first oscillator
$x$ is subject to a friction force $\mu\dot{x}$ ($\mu>0$), while the second
oscillator $y$ is subject to an antifriction force $-\nu\dot{y}$ ($\nu>0$). The
parameters $\mu$ and $\nu$ are a measure of the loss and gain. The oscillators
are coupled linearly and the coupling strength is represented by the parameter
$\epsilon$. The equations of motion of these oscillators are
\begin{eqnarray}
\ddot{x}+\omega^2x+\mu\dot{x}&=&-\epsilon y,\nonumber\\
\ddot{y}+\omega^2y-\nu\dot{y}&=&-\epsilon x.
\label{E1}
\end{eqnarray}

To treat this system at a classical level, we seek solutions to (\ref{E1}) of
the form $e^{i\lambda t}$. The frequency $\lambda$ satisfies the quartic
polynomial equation 
\begin{eqnarray}
&&\lambda^4-i(\mu-\nu)\lambda^3-(2\omega^2-\mu\nu)\lambda^2\nonumber\\
&&\qquad\qquad\qquad+i\omega^2(\mu-\nu)\lambda-\epsilon^2+\omega^4=0.
\label{E2}
\end{eqnarray}

An important special case arises when the loss and gain are balanced; that is,
when $2\gamma=\mu=\nu$. In this case the frequencies $\lambda$ are the roots
of the quartic polynomial $f(\lambda)$, where
\begin{equation}
f(\lambda)=\lambda^4-(2\omega^2-4\gamma^2)\lambda^2-\epsilon^2+\omega^4.
\label{E3}
\end{equation}
For this special case the classical equations of motion (\ref{E1}) can be
derived from the Hamiltonian \cite{R13}
\begin{equation}
H=pq+\gamma(yq-xp)+\left(\omega^2-\gamma^2\right)xy+\epsilon\left(x^2+y^2\right)
/2.
\label{E4}
\end{equation}

If the coupling parameter $\epsilon$ of the $x$ and $y$ oscillators is set to
zero, this Hamiltonian reduces to the Hamiltonian considered by Bateman
\cite{R14}. In his paper Bateman sought a variational principle to derive an
equation of motion having a friction term linear in velocity. To do so, he
introduced an additional degree of freedom; namely, a time-reversed version of
the original damped harmonic oscillator. This auxiliary oscillator acts as an
energy reservoir and can be considered as an effective description of a thermal
bath. The classical Hamiltonian for the Bateman system was constructed by Morse
and Feschbach \cite{R15} and the corresponding quantum theory was analyzed by
many authors, including Bopp \cite{R16}, Feshbach and Tikochinsky \cite{R17},
Tikochinsky \cite{R18}, Dekker \cite{R19}, Celeghini, Rasetti, and Vitiello
\cite{R20}, Banerjee and Mukherjee \cite{R21}, and Chru\'sci\'nski and Jurkowski
\cite{R24}. We emphasize that in all these references only the {\it
noninteracting} ($\epsilon=0$) case was considered. It is easy to see that the
Hamiltonian (\ref{E4}) is $\cPT$ symmetric, where the action of parity $\cP$ is
to interchange the loss and gain oscillators and its effect is given by
\cite{R22}
\begin{equation}
\cP:~x\to-y,\quad y\to-x,\quad p\to-q,\quad q\to-p,
\label{E5}
\end{equation}
whilst the action of time reversal $\cT$ is to change the signs of the momenta
\begin{equation}
\cT:~x\to x,\quad y\to y,\quad p\to-p,\quad q\to-q.
\label{E6}
\end{equation}
Note that $H$ is {\it not} symmetric under $\cP$ or $\cT$ separately, but it is
symmetric under combined $\cP$ and $\cT$. For a one-dimensional system, $\cP$
reduces to the usual parity operator $\cP:\,x\to-x,~p\to-p$ and $\cT$ is the
usual time-reversal operator.
Because the balanced-loss-and-gain system is Hamiltonian, the energy (the value
of $H$) is conserved; that is, its numerical value is constant in time. However,
the expression for the energy in (\ref{E4}) is not recognizable as a simple sum
of kinetic and potential energy (such as $p^2+q^2+x^2+y^2$).

The noteworthy feature of $\cPT$-symmetric systems with balanced loss and gain
is that they exhibit phase transitions. When the coupling of the two oscillators
is small, the energy flowing into the $y$ resonator
cannot transfer fast enough to the $x$ resonator, where the energy is flowing
out. Thus, the system cannot be in equilibrium. However, when the coupling
constant $\epsilon$ exceeds a critical value, all of the energy flowing into the
$y$ resonator can transfer to the $x$ resonator and the entire system can attain
equilibrium. The system is in equilibrium only if the frequencies are real
because complex frequencies indicate that there is exponential growth and decay.

\begin{widetext}
To understand why there are phase transitions we plot the quartic polynomial
$f(\lambda)$ in (\ref{E3}) to see whether this polynomial cuts the horizontal
axis in four places (in which case there are four real frequencies), two places
(in which case there are two real frequencies and two complex frequencies), or
not at all (here, there are four complex frequencies). As one can see in
Fig.~\ref{F1}, for small values of $\epsilon$ there are no real frequencies,
but as $\epsilon$ increases there is a transition at $\epsilon_1=2\gamma\sqrt{
\omega^2-\gamma^2}$ to a situation where there are four real frequencies.
Interestingly, one can see that when the coupling $\epsilon$ is sufficiently
large, there is a second transition at $\epsilon_2=\omega^2$. This transition
is difficult to see in classical experiments because in the strong-coupling
regime the loss and gain components must be so close that they overlap and
therefore interfere with one another. For example, in the pendulum experiment in
Ref.~\cite{R11} the pendula would have to be so close that they touch and could
no longer swing freely. This strong-coupling region is discussed for the case of
coupled systems {\it without} loss and gain in Ref.~\cite{R23}, where it is 
referred to as the {\it ultrastrong-coupling regime}. 

\begin{figure}[h!]
\begin{center}
\includegraphics[scale=0.20]{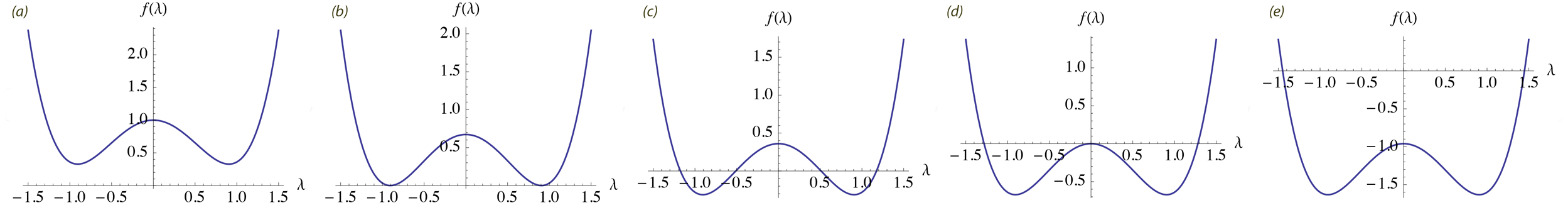}
\end{center}
\caption{Five plots of $f(\lambda)$ in (\ref{E3}) for $\lambda$ in the range
$-1.5<\lambda<1.5$. In these plots $\omega=1.0$, $\gamma=0.3$, and $\epsilon$ 
has the values (a) $\epsilon=0.01$ (this value of $\epsilon$ lies in the first
broken $\cPT$ region); (b) $\epsilon=\epsilon_1=2\gamma\sqrt{\omega^2-\gamma^2}
\approx0.572364$ (this is the first transition); (c) $\epsilon=0.8$ (this value
of $\epsilon$ lies in the unbroken $\cPT$ region in which the frequencies are
all real); (d) $\epsilon=\epsilon_2=\omega^2=1.0$ (this is the location of the
second transition); (e) $\epsilon=1.4$ (this value of $\epsilon$ lies in the
second broken $\cPT$ region).}
\label{F1}
\end{figure} 

The signal that the system is in equilibrium is that the resonators exhibit {\it
Rabi oscillations} (power oscillations between the two resonators) as shown in
Fig.~\ref{F2}. Note that the Rabi oscillations are $90^\circ$ out of phase.

\begin{figure}[h!]
\begin{center}
\includegraphics[scale=0.20]{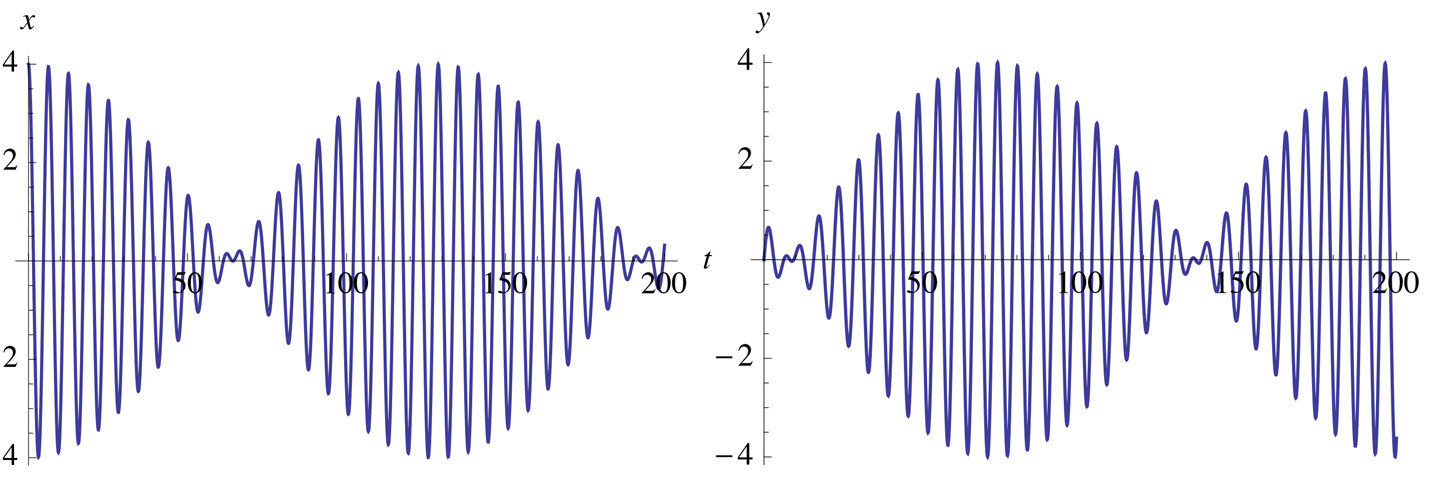}
\end{center}
\caption{Rabi oscillations in the unbroken $\cPT$-symmetric region. In this
figure $\gamma=0.01$, $\epsilon=0.05$, and $\omega=1.0$.}
\label{F2}
\end{figure}

This paper is organized as follows: In Sec.~\ref{s3} we examine the classical
solutions to (\ref{E1}). Next, in Sec.~\ref{s4} we examine the quantized version
of the system described by the Hamiltonian (\ref{E4}). We identify the quantum
analogs of the $\cPT$ phase transitions and show that the quantum and classical
transitions occur at exactly the same values of the physical parameters.
Finally, in Sec.~\ref{s5} we make some brief concluding remarks.

\section{Classical interpretation}
\label{s3}

\subsection{Balanced loss and gain}
\label{ss3a}

When $2\gamma=\mu=\nu$, the quartic equation (\ref{E2}) for $\lambda$ reduces to
the biquadratic equation (\ref{E3}) whose solutions are
\begin{equation}
\lambda^2=\omega^2-2\gamma^2\pm2\sqrt{\epsilon^2-4\gamma^2\omega^2+4\gamma^4}.
\label{E7}
\end{equation}
There are four real frequencies $\lambda$ when $\epsilon$ is in the range
\begin{equation}
2\gamma\sqrt{\omega^2-\gamma^2}<\epsilon<\omega^2.
\label{E8}
\end{equation}
This is the unbroken classical $\cPT$-symmetric region.

We plot the real and imaginary parts of the frequency $\lambda$ in Fig.~\ref{F3}
for the values $\omega=1.0$ and $\gamma=0.01$. For these parametric values the
$\cPT$ phase transition occurs at $\epsilon_1=2\gamma\sqrt{\omega^2-\gamma^2}
\approx0.019999$. When $\epsilon$ is below this critical value, the real part of
$\lambda$ has one positive value, which is shown in Fig.~\ref{F3}, and one
negative value. Also, below the critical value, the imaginary part of $\lambda$
is nonzero, as shown in Fig.~\ref{F3}. As $\epsilon$ approaches the critical
value from below, the imaginary part of $\lambda$ vanishes and the real part of
$\lambda$ bifurcates.

\begin{figure}[b!]
\begin{center}
\includegraphics[scale=0.20]{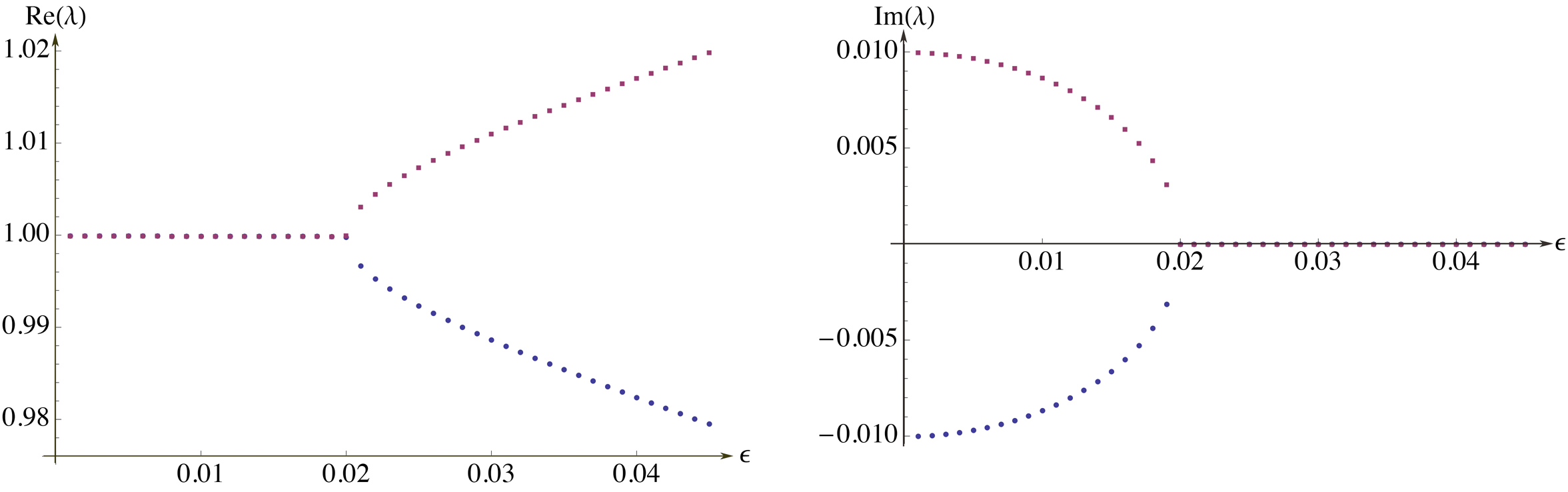}
\end{center}
\caption{A plot of the real and imaginary parts of the classical frequency
$\lambda$ in (\ref{E7}) for $\epsilon$ near the $\cPT$ phase transition at
$\epsilon=\epsilon_1\approx0.019999$. For this figure $\omega=1.0$ and $\gamma=
0.01$. Note that at the phase transition the real and imaginary parts of the
frequency bifurcate in an orthogonal direction.}
\label{F3}
\end{figure}

A second transition occurs when $\epsilon=\epsilon_2=\omega^2=1$. Above this
transition point there is now only one super mode, as shown in Fig.~\ref{F4},
instead of two pairs of real frequencies.

\begin{figure}[h!]
\begin{center}
\includegraphics[scale=0.20]{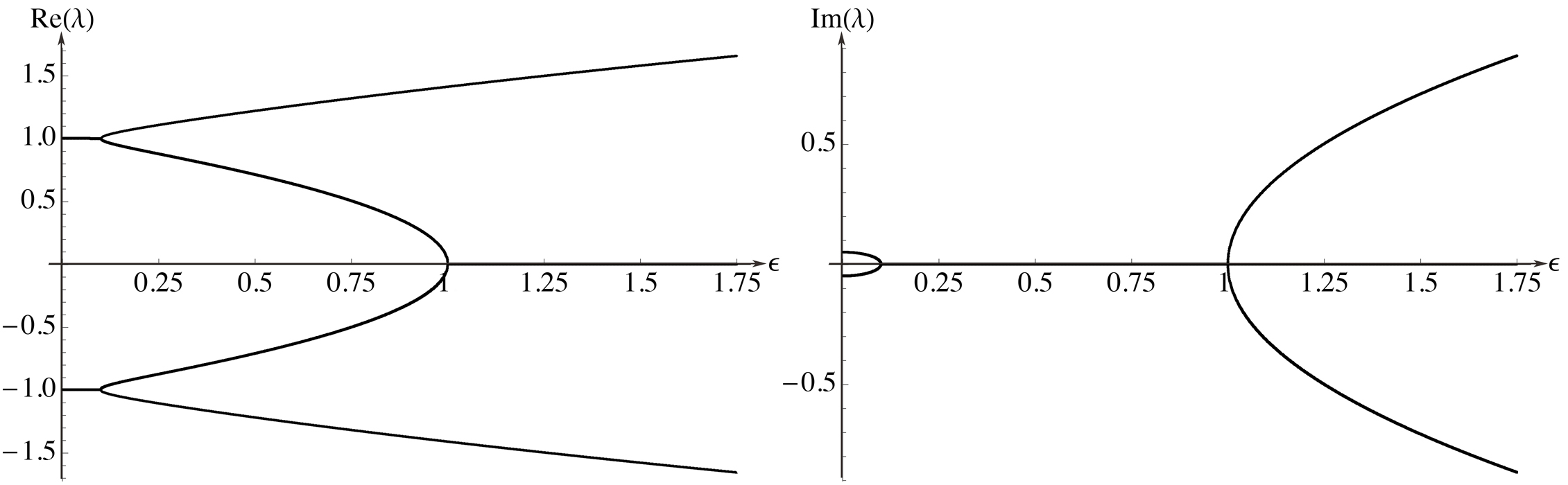}
\end{center}
\caption{A plot of the real and imaginary parts of $\lambda$ for $0\leq\epsilon
\leq1.75$. Observe that there is a second transition at $\epsilon=\epsilon_2=
\omega^2=1$. For this figure $\omega=1.0$ and $\gamma=0.01$.}
\label{F4}
\end{figure}

\subsection{Unbalanced loss and gain}
\label{ss3b}

Let us consider the general case (\ref{E2}) in which $\mu\neq\nu$ (that is, the
loss and gain are not exactly balanced. In this case the sharp transition from a
region of broken $\cPT$ symmetry to a region of an unbroken symmetry disappears
and there is only an approximate transition. To see this approximate transition,
we take $\mu=0.04$ and $\nu=0.01$ and plot the classical frequencies $\lambda$
in Fig.~\ref{F5}. In contrast with Fig.~\ref{F3}, the frequency $\lambda$ is
never exactly real. Rather, there is one region of $\epsilon$ in which the
difference between the imaginary parts of the frequencies is big and the
difference between the real parts of the frequencies is small but {\it nonzero},
and a second region in which the difference between the imaginary parts of the
frequencies is small but {\it nonzero} and the difference between the real parts
of the frequencies is big. Unlike the behavior shown in Fig.~\ref{F3}, at the
approximate transition in Fig.~\ref{F5} the real parts of the frequencies do
not separate in an orthogonal direction but rather separate smoothly.

\begin{figure}[h!]
\begin{center}
\includegraphics[scale=0.20]{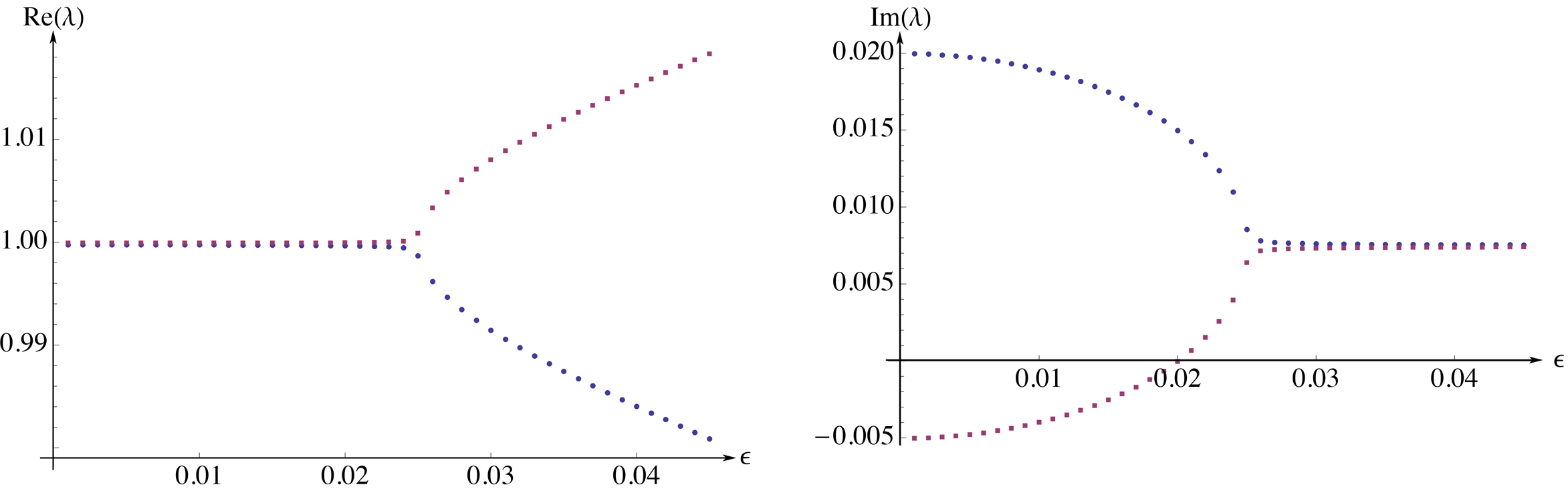}
\end{center}
\caption{Real and imaginary parts of the classical frequencies $\lambda$ for the
unbalanced case in which the loss and gain parameters $\mu$ and $\nu$ are
unequal. Compared with Fig.~\ref{F3}, the frequencies separate smoothly as a
function of $\epsilon$ at the approximate transition point. For this figure
$\mu=0.04$, $\nu=0.01$, and $\omega=1.0$. It is virtually impossible to have a
physical system in which the loss and gain are exactly balanced, so this figure
should be regarded as physically realistic while Fig.~\ref{F3} is an
idealization.}
\label{F5}
\end{figure}

\end{widetext}

We can treat this problem perturbatively by taking $\mu$, $\nu$, and $\epsilon$
small compared with the natural frequency $\omega$. We let $\mu=\alpha\epsilon$
and $\nu=\beta\epsilon$ and expand $\lambda$ in (\ref{E2}) in powers of the
small parameter $\epsilon$:
$\lambda=\lambda_0(1+\epsilon\lambda_1+\epsilon^2\lambda_2+\ldots)$.
To zeroth order $\lambda_0=\pm i\omega$. To first order our results are
consistent with the plots in Fig.~\ref{F5}: For $\epsilon>(\mu+\nu)\omega$,
\begin{equation}
\lambda=\left\{
\begin{array}{l}
i\,\omega\pm i\,\epsilon\frac{\sqrt{4-(\alpha+\beta)^2\omega^2}}{4\omega}
+\epsilon \frac{(\alpha-\beta)}{4},\\ \\
-i\,\omega\pm i\,\epsilon\frac{\sqrt{4-(\alpha+\beta)^2\omega^2}}{4\omega}+
\epsilon \frac{(\alpha-\beta)}{4}.\\ \end{array}\right.
\label{E10}
\end{equation}
and for $\epsilon<(\mu+\nu)\omega$,
\begin{equation}
\lambda= \left\{ \begin{array}{l}
i\,\omega\pm\epsilon\frac{\sqrt{(\alpha+\beta)^2\omega^2-4}}{4\omega}+\epsilon
\frac{(\alpha-\beta)}{4},\\ \\
-i\,\omega\pm\epsilon\frac{\sqrt{(\alpha+\beta)^2\omega^2-4}}{4\omega}+\epsilon
\frac{(\alpha-\beta)}{4}.\\
\label{E11}
\end{array}
\right.
\end{equation}

\section{Quantum interpretation}
\label{s4}

When the loss and gain parameters in (\ref{E1}) are equal, the coupled
oscillator system is described by the Hamiltonian $H$ in (\ref{E4}). To quantize
this classical Hamiltonian we replace the classical variables $p$, $q$, $x$, and
$y$ with the corresponding quantum operators that satisfy the commutator
equations $[x,p]=[y,q]=i$ and $[x,y]=[p,q]=[x,q]=[y,p]=0$. In Subsec.~\ref{ss4a}
we discuss the eigenfunctions of $H$ and in Subsec.~\ref{ss4b} we discuss the
eigenvalues.

\subsection{Eigenfunctions of $H$}
\label{ss4a}
The eigenfunctions of the Hamiltonian (\ref{E4}) satisfy the
time-independent Schr\"odinger equation
\begin{eqnarray} 
&&\left[-\partial_x \partial_y-i\gamma(y\partial_y -
x\partial_x)+(\omega^2-\gamma^2)xy\right] \psi_{m,n}(x,y)\nonumber\\
&&+\frac{\epsilon}{2}\left(x^2+y^2\right)\psi_{m,n}(x,y)=E_{m,n}\psi_{m,n}(x,y).
\label{E12}
\end{eqnarray} 
The eigenvalues $E_{m,n}$ correspond to the eigenfunctions $\psi_{m,n}(x,y)$.
The eigenfunctions have the general form
\begin{equation}
\psi_{m,n}(x,y)=e^{-(2axy+bx^2+cy^2)/2}P_{m,n}(x,y),
\label{E13}
\end{equation}
where
\begin{equation} 
b=c^*=\frac{\epsilon}{2(a+i\gamma)}
\label{E14}
\end{equation} 
and $a$ is a solution to the quartic equation $g(a)=0$, where
\begin{equation}
g(a)=a^4+(2\gamma^2-\omega^2)a^2+\epsilon^2/4+\gamma^4-\gamma^2\omega^2.
\label{E15}
\end{equation}
The quantities $P_{m,n}(x,y)$ are polynomials in $x$ and $y$. The index $n$ is a
nonnegative integer $(n=0,1,2,3,\ldots)$ while the index $m$ is an integer that
runs from $0$ to $n$. Thus, the polynomials form a Pascal-like triangle in which
the first index $m$ labels the row and $n$ labels the column:
\begin{equation}\begin{array}{ccccccccccc}
&& & & &P_{0,0}& & & && \\
&& & &P_{1,0}& &P_{1,1}& & && \\
&& &P_{2,0}& &P_{2,1}& &P_{2,2}& && \\
&&P_{3,0}& &P_{3,1}& &P_{3,2}& &P_{3,3}&& \\
&P_{4,0}& &P_{4,1}& &P_{4,2}& &P_{4,3}& &P_{4,4}&\\
\Ddots& &\Ddots& &\Ddots& &\ddots& &\ddots& &\ddots \\
\end{array}
\nonumber
\end{equation}

In terms of the quantity
\begin{equation}
\Delta=\sqrt{4bc-\gamma^2}
\label{E16}
\end{equation}
the first seven polynomials are
\begin{eqnarray}
P_{0,0}&=&1,\nonumber\\
P_{1,0}&=&\frac{i\gamma-\Delta}{2c}\,x+y,\nonumber\\
P_{1,1}&=&\frac{\Delta+i\gamma}{2c}x+y,\nonumber\\
P_{2,0}&=&\left[\frac{(i\gamma-\Delta)x}{2c}+y\right]^2
-\frac{i\gamma-\Delta}{c(2a-\Delta)},\nonumber\\
P_{2,1}&=&\left[\frac{i\gamma x}{2c}+y\right]^2-\frac{\Delta^2x^2}{4c^2}
-\frac{i\gamma}{2ac},\nonumber\\
P_{2,2}&=&\left[\frac{(\Delta+i\gamma)x}{2c}+y\right]^2
-\frac{\Delta+i\gamma}{c(2a+\Delta)},\nonumber\\
P_{3,0}&=&\left[\frac{(i\gamma-\Delta)x}{2c}+y\right]^3\nonumber\\
&&-\frac{3(i\gamma-\Delta)}{2c(2a-\Delta)}
\left[\frac{(i\gamma-\Delta)x}{2c}+y\right].
\label{E17}
\end{eqnarray}

The polynomials $P_{m,n}$ satisfy satisfy two three-term recursion relations,
one in the first index with the second index held fixed at $0$ (at the left edge
of the Pascal triangle),
\begin{eqnarray}
P_{n+1,0}=\frac{(i\gamma-\Delta)x+2cy}{2c}P_{n,0}
+\frac{n(\Delta-i\gamma)}{c(2a-\Delta)}P_{n-1,0},
\label{E18}
\end{eqnarray}
and another with both indices being equal (at the right edge of the Pascal
triangle),
\begin{eqnarray}
P_{n+1,n+1}&=&\frac{(\Delta+i\gamma)x+2cy}{2c}P_{n,n}\nonumber\\
&&\qquad-\frac{n(\Delta+i\gamma)}{c(2a+\Delta)}P_{n-1,n-1}.
\label{E19}
\end{eqnarray}

The operators $\partial_x$ and $\partial_y$ are lowering operators for
the polynomials $P_{n,0}$ and $P_{n,n}$:
\begin{eqnarray}
&&\!\!\!\!\!\!\!\!\!\!\!\!\!\!\!\!\partial_xP_{n,0}=n\,\frac{-\Delta+i\gamma}
{2c}P_{n-1,0},~~\partial_yP_{n,0}=n\,P_{n-1,0},\nonumber\\
&&\!\!\!\!\!\!\!\!\!\!\!\!\!\!\!\!\partial_xP_{n,n}=n\,\frac{\Delta+i\gamma}
{2c}P_{n-1,n-1},~~\partial_yP_{n,n}=n\,P_{n-1,n-1}.
\label{E20}
\end{eqnarray}
These equations are the analogs of the relation $\partial_x H_n(x)=nH_{n-1}(x)$
for the Hermite polynomials $H_n(x)$. 

Upon substituting (\ref{E20}) into (\ref{E18}) and (\ref{E19}), we obtain the
relations
\begin{eqnarray}
P_{n,0}&=&\frac{i\gamma-\Delta}{2c}\,x\,P_{n-1,0}+y\,P_{n-1,0}\nonumber\\
&&\quad+\frac{\Delta-i\gamma}
{c(2a-\Delta)}\,\partial_yP_{n-1,0}
\label{E21}
\end{eqnarray}
and
\begin{eqnarray}
P_{n,n}&=&\frac{\Delta+i\gamma}{2c}\,x\,P_{n-1,n-1}+y\,P_{n-1,n-1}\nonumber\\
&&\quad-\frac{2}{2a+\Delta}\,\partial_xP_{n-1,n-1},
\label{E22}
\end{eqnarray}
from which we obtain the differential equation satisfied by the polynomials
at the left and right edges of the Pascal triangle:
\begin{eqnarray}
\left[\frac{(i\gamma-\Delta)x+2cy}{2c}\partial_y+\frac{\Delta-i\gamma}
{c(2a-\Delta)}\partial_y^2\right]P_{n,0}=n\,P_{n,0},
\label{E23}
\end{eqnarray}
\begin{eqnarray}
\left[\frac{2bx+(\Delta-i\gamma)y}{2b}\partial_x+\frac{\Delta+i\gamma}
{c(2a-\Delta)}\partial_x^2\right]P_{n,n}=n\,P_{n,n}.
\label{E24}
\end{eqnarray}

We can also construct operators that connect the polynomials on a given
horizontal level in the Pascal triangle. To do this we define the {\it left
shift operator} ${\bf L}$ as
\begin{equation}
{\bf L}\equiv(y+r_1 x)\partial_y-(r_1^*y+x)\partial_x+r_2\partial_y^2-r_2^*
\partial_x^2+r_3\partial_x\partial_y,
\label{E25}
\end{equation}
where
\begin{eqnarray}
r_1&=&\frac{i\gamma-\Delta}{2c},\nonumber\\
r_2&=&\frac{(2a+i\gamma)(\Delta-i\gamma)}{4ac(2a-\Delta)},\nonumber\\
r_3&=&\frac{i\gamma}{a(2a-\Delta)}.
\label{E26}
\end{eqnarray}
The effect of ${\bf L}$ on $P_{m,n}$ is
$$ {\bf L}\,P_{m,n}=n\frac{\Delta^2+i\Delta\gamma}{2bc}\,P_{m,n-1}.$$
We also define the {\it right shift operator} ${\bf R}$ as
\begin{equation}
{\bf R}\equiv(x+s_1 y)\partial_x-(s_1^*x+y)\partial_y+s_2\partial_x^2-s_2^*
\partial_y^2+s_3\partial_x\partial_y,
\label{E27}
\end{equation}
where
\begin{eqnarray}
s_1&=&\frac{\Delta-i\gamma}{2c},\nonumber\\
s_2&=&-\frac{(2a-i\gamma)(\Delta-i\gamma)}
{4ab(2a+\Delta)},\nonumber\\
s_3&=&-\frac{i\gamma}{a(2a+\Delta)}.
\label{E28}
\end{eqnarray}
The effect of ${\bf R}$ on $P_{m,n}$ is
$${\bf R}\,P_{m,n}=(m-n)\,\frac{-\Delta^2+i\Delta\gamma}{2bc}\,P_{m,n+1}.$$

Note that $P_{n,m}$ are eigenstates of the operators ${\bf LR}$ and ${\bf RL}$:
\begin{equation}
{\bf LR}\,P_{m,n}=(n-1)(n-m)\,\frac{\Delta^2}{bc}\,P_{m,n}
\label{E29}
\end{equation}
for $n=0,1,\ldots,m$, and
\begin{equation}
{\bf RL}\,P_{m,n}=n(n-m-1)\,\frac{\Delta^2}{bc}\,P_{m,n}
\label{E30}
\end{equation}
for $n=0,1,\ldots,m$. If we combine (\ref{E29}) and (\ref{E30}), we obtain the
interesting result
\begin{equation}
[{\bf R},{\bf L}]\,P_{m,n}=-m\,\frac{\Delta^2}{bc}\,P_{m,n}.
\label{E31}
\end{equation}

\subsection{Eigenvalues of $H$}
\label{ss4b}
The eigenvalues have the general form
\begin{equation}
E_{m,n}=(m+1)a+(n-m/2)\Delta,
\label{E32}
\end{equation}
where $m=0,1,2,\ldots$ and $n=0,1,2,\ldots m$. Note that there are four possible
spectra of eigenvalues corresponding to the four possible solutions for $a$,
which are the roots of $g(a)$ in (\ref{E15}):
\begin{eqnarray} 
a_1&=&-\frac{1}{2}\sqrt{2\omega^2-4\gamma^2-2\sqrt{\omega^4-\epsilon^2}},
\nonumber\\
a_2&=&\frac{1}{2}\sqrt{2\omega^2-4\gamma^2-2\sqrt{\omega^4-\epsilon^2}},
\nonumber\\
a_3&=&-\frac{1}{2}\sqrt{2\omega^2-4\gamma^2+2\sqrt{\omega^4-\epsilon^2}},
\nonumber\\
a_4&=&\frac{1}{2}\sqrt{2\omega^2-4\gamma^2+2\sqrt{\omega^4-\epsilon^2}}.
\label{E33}
\end{eqnarray} 
Although there are four possible sets of eigenvalues, we will see that only
one of these sets is physically acceptable; that is, there is only one set of
eigenvalues that is real and bounded below. This set only occurs in the {\it
classical unbroken} $\cPT$-symmetric region of $\epsilon$ in (\ref{E8}).

\begin{widetext}
Corresponding to the two classical phase transitions discussed earlier, there
are also two quantum transitions at the same values of the coupling $\epsilon_1$
and $\epsilon_2$ as the classical transitions. To locate the quantum phase
transitions we plot in Fig.~\ref{F6} the quartic polynomial $g(a)$ in
(\ref{E15}) as a function of $a$ for various values of $\epsilon$ and observe
whether this polynomial cuts the horizontal axis in four places, two places, or
not at all.
\begin{figure}[h!]
\begin{center}
\includegraphics[scale=0.20]{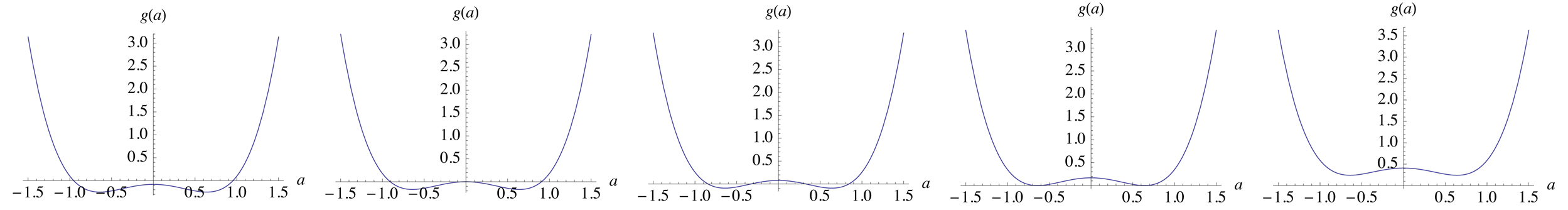}
\end{center}
\caption{A plot of $g(a)$ in (\ref{E15}) for $-1.5<a<1.5$, $\omega=1.0$, $\gamma
=0.3$, where (a) $\epsilon=0.01$ (this is in the first broken region), (b)
$\epsilon=\epsilon_1=2\gamma\sqrt{\omega^2-\gamma^2}\approx 0.572364$ (this is
the location of the first transition); (c) $\epsilon=0.8$ (this is in the
unbroken-$\cPT$ region in which the classical frequencies are all real); (d)
$\epsilon=\epsilon_2=\omega^2=1.0$ (this is the location of the second
transition); (e) $\epsilon=1.4$ (this is in the second broken $\cPT$ region).}
\label{F6}
\end{figure} 
\end{widetext}

It is important to understand why there are four possible sets of
quantum eigenvalues. This comes about because there are four possible pairs
of Stokes wedges in the complex domain in which the eigenfunctions $\psi$ in
(\ref{E13}) vanish exponentially. To explain what is going on, we use, as an
elementary example, the quantum harmonic oscillator, whose Hamiltonian is
$H=p^2+x^2$. One set of eigenfunctions $\psi$ of this Hamiltonian in complex-$x$
space have the form $\psi_n(x)=e^{-x^2/2}P_n(x)$, where $P_n(x)$ is a Hermite
polynomial. These eigenfunctions vanish exponentially in a pair of Stokes wedges
of opening angle $\pi/2$ centered about the positive and negative real axes in
the complex-$x$ plane. The eigenvalues $E_n=2n+1$ ($n=0,\,1,\,2,\,\ldots$)
associated with these eigenfunctions are real and bounded below. There is a
second set of eigenfunctions of the form $\psi_n(x)=e^{x^2/2}P_n(x)$, where
$P_n(x)$ is again a Hermite polynomial. These eigenfunctions vanish
exponentially in a pair of Stokes wedges of opening angle $\pi/2$ centered about
the positive and negative {\it imaginary} axes in the complex-$x$ plane. The
eigenvalues $E_n=-2n-1$ ($n=0,\,1,\,2,\, \ldots$) associated with these
eigenfunctions are real and bounded above. A full description of these two
classes of eigenfunctions and eigenvalues is given in Ref.~\cite{A1}.

For the coupled-oscillator problem discussed in this paper the eigenfunctions
have the general form (\ref{E13}). The exponential component of these
eigenfunctions can be rewritten as
\begin{equation}
e^{-(2axy+bx^2+cy^2)/2}=e^{-(bu^2+Ry^2)/2},
\label{E34}
\end{equation}
where
\begin{equation}
u=x+ay/b\quad{\rm and}\quad R=c-a^2/b.
\label{E35}
\end{equation}
It is important to determine the Stokes wedges in the complex-$u$ plane and in
the complex-$y$ plane in which the eigenfunctions vanish. We consider each of
the four values of $a$ in (\ref{E33}) in turn.

\begin{figure}[h!]
\begin{center}
\includegraphics[scale=0.24]{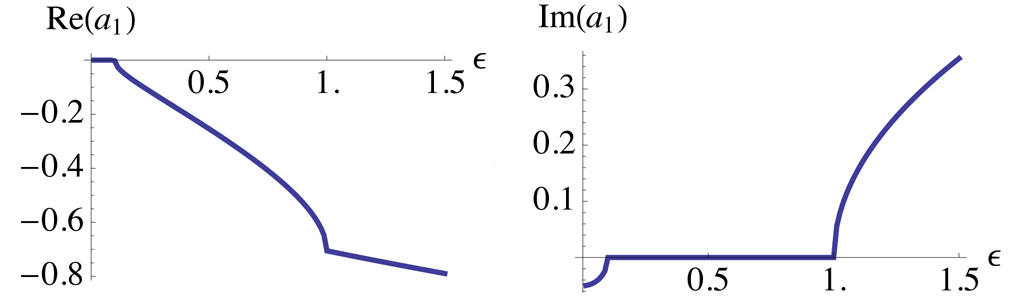}
\end{center}
\caption{Real and imaginary parts of $a_1$ in (\ref{E33}) plotted as functions
of $\epsilon$ for $0\leq\epsilon\leq1.5$. For this plot we have taken $\gamma=
0.05$ and $\omega=1.0$. For these values the region of unbroken-$\cPT$ symmetry
is $0.0998749\leq\epsilon\leq1.0$.}
\label{F7}
\end{figure}

\begin{figure}[h!]
\begin{center}
\includegraphics[scale=0.24]{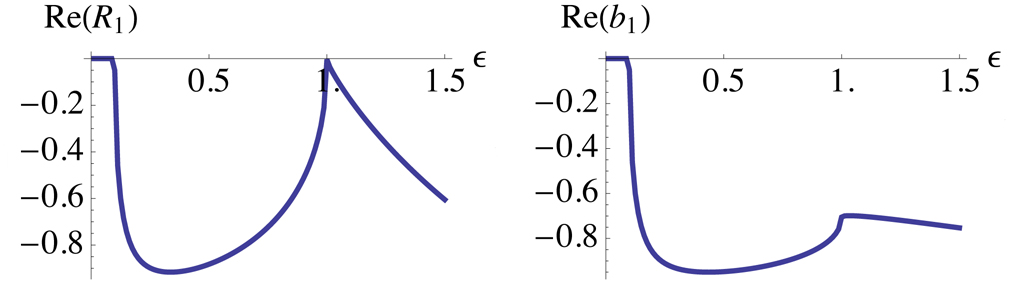}
\end{center}
\caption{Plots of the real parts of $R_1$ and $b_1$ (the values of $R$ and
$b$ corresponding to $a=a_1$) for $0\leq\epsilon\leq1.5$. For this plot $\gamma=
0.05$ and $\omega=1.0$.}
\label{F8}
\end{figure}

First, we consider $a_1$ in (\ref{E33}). In Fig.~\ref{F7} we plot the real and
imaginary parts of $a_1$ as functions of $\epsilon$ and see that $a_1$ is real
in the unbroken-$\cPT$ region (\ref{E8}). Furthermore, $\Delta$ in (\ref{E16})
is real and positive in this region. Thus, the eigenvalues in (\ref{E32}) are
real.  
In the unbroken region ${\rm Re}\,b$ in (\ref{E14}) and ${\rm Re}\,R$ in
(\ref{E35}) are both negative (see Fig.~\ref{F8}). Thus, the eigenfunctions
vanish exponentially in pairs of $90^\circ$-Stokes wedges centered about the
imaginary axes in the $u$ and $y$ planes.
However, since $a_1$ is negative, the eigenspectrum (\ref{E32}) is not bounded
below, and thus this case must be rejected on physical grounds.

\begin{figure}[h!]
\begin{center}
\includegraphics[scale=0.24]{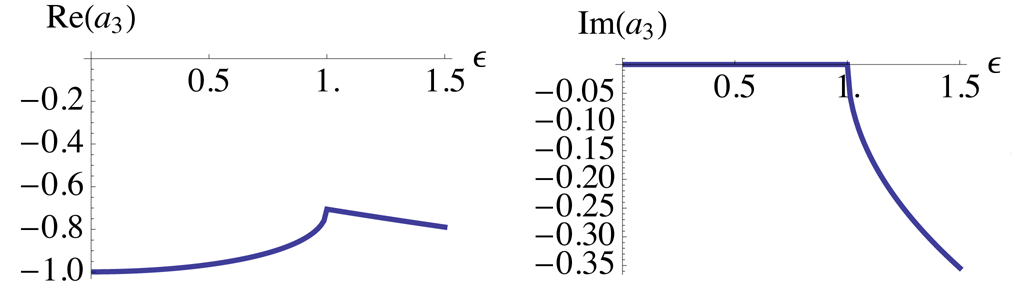}
\end{center}
\caption{Real and imaginary parts of $a_3$ in (\ref{E33}) plotted as functions
of $\epsilon$ for $0\leq\epsilon\leq1.5$. For this plot we have taken $\gamma=
0.05$ and $\omega=1.0$.}
\label{F9}
\end{figure}

\begin{figure}[h!]
\begin{center}
\includegraphics[scale=0.24]{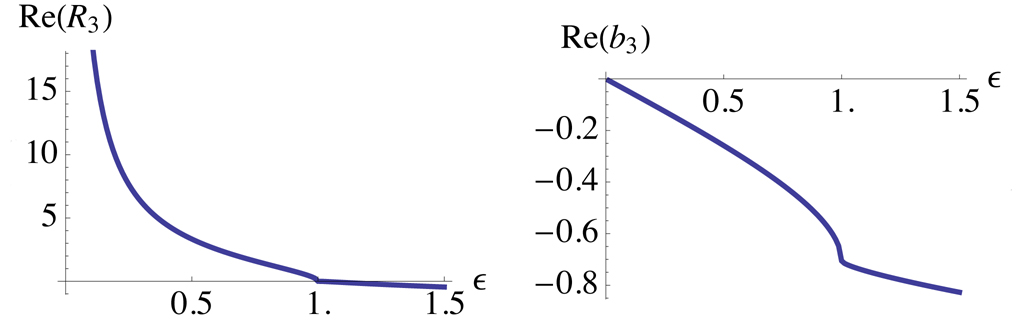}
\end{center}
\caption{Plots of the real parts of $R_3$ and $b_3$ for $0\leq\epsilon\leq1.5$.
For this plot $\gamma=0.05$ and $\omega=1.0$.}
\label{F10}
\end{figure}

Next, we consider $a_3$ in (\ref{E33}). In Fig.~\ref{F9} we plot $a_3$ as a
function of $\epsilon$ and see that $a_3$ is real in both the first
broken-$\cPT$ region and the unbroken-$\cPT$ region (\ref{E8}). Furthermore,
$\Delta$ in (\ref{E16}) is real and positive in the unbroken region of
$\epsilon$. Thus, the eigenvalues in (\ref{E32}) are real in the unbroken-$\cPT$
region.  
In the unbroken region ${\rm Re}\,b$ in (\ref{E14}) is negative and ${\rm Re}\,
R$ in (\ref{E35}) is positive (see Fig.~\ref{F10}). Thus, the eigenfunctions
vanish exponentially in pairs of $90^\circ$-Stokes wedges centered about the
imaginary axis in the $u$ plane and centered about the real axis in the $y$
plane.
However, $a_3$ is negative, so the eigenspectrum (\ref{E32}) is not bounded
below, and again this case must be rejected on physical grounds.

\begin{figure}[h!]
\begin{center}
\includegraphics[scale=0.24]{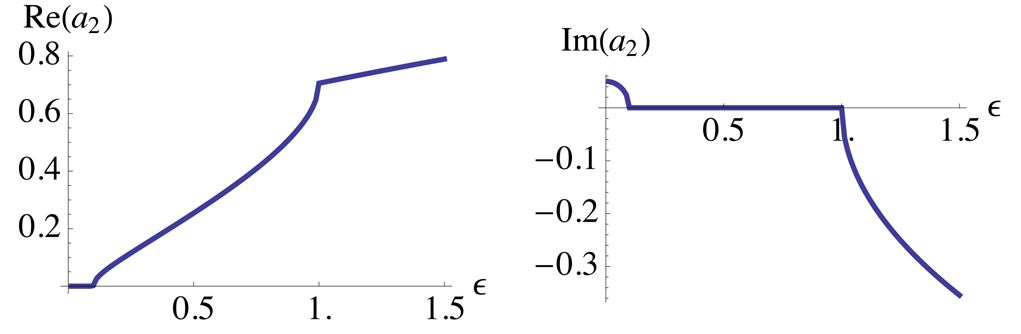}
\end{center}
\caption{Real and imaginary parts of $a_2$ in (\ref{E33}) plotted as functions
of $\epsilon$ for $0\leq\epsilon\leq1.5$. For this plot we have taken $\gamma=
0.05$ and $\omega=1.0$.}
\label{F11}
\end{figure}

\begin{figure}[h!]
\begin{center}
\includegraphics[scale=0.24]{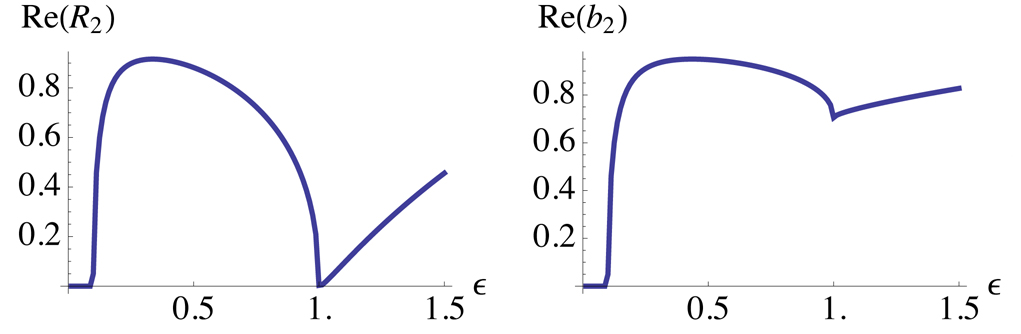}
\end{center}
\caption{Plots of the real parts of $R_2$ and $b_2$ for $0\leq\epsilon\leq1.5$.
For this plot $\gamma=0.05$ and $\omega=1.0$.}
\label{F12}
\end{figure}

Next, we consider $a_2$ in (\ref{E33}). In Fig.~\ref{F11} we plot $a_2$ as a
function of $\epsilon$ and see that $a_2$ is real and positive in the
unbroken-$\cPT$ region (\ref{E8}). Again, $\Delta$ in (\ref{E16}) is real and
positive in this region. Thus, the eigenvalues in (\ref{E32}) are real and
positive.  
In the unbroken region ${\rm Re}\,b$ in (\ref{E14}) and ${\rm Re}\,R$ in
(\ref{E35}) are both positive (see Fig.~\ref{F12}). Thus, the eigenfunctions
vanish exponentially in pairs of $90^\circ$-Stokes wedges centered about the
real axes in both the $u$ and $y$ planes.
Because the eigenspectrum (\ref{E32}) is bounded below and the eigenfunctions
vanish exponentially in the appropriate Stokes wedges, we regard this as a
physically acceptable case.

\begin{figure}[h!]
\begin{center}
\includegraphics[scale=0.24]{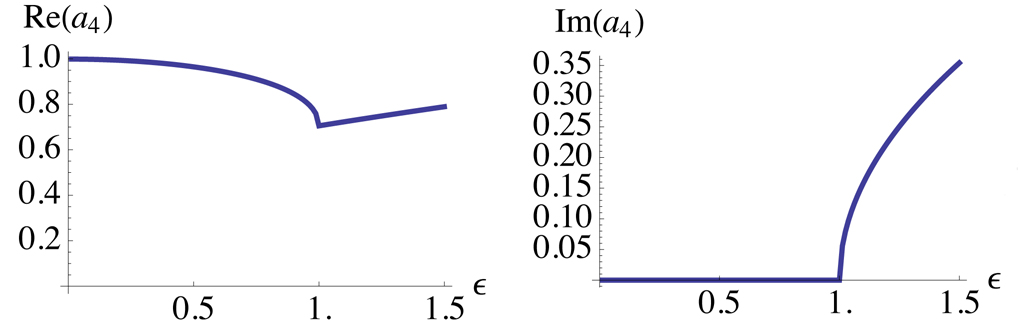}
\end{center}
\caption{Real and imaginary parts of $a_4$ in (\ref{E33}) plotted as functions
of $\epsilon$ for $0\leq\epsilon\leq1.5$. For this plot we have taken $\gamma=
0.05$ and $\omega=1.0$.}
\label{F13}
\end{figure}

\begin{figure}[h!]
\begin{center}
\includegraphics[scale=0.24]{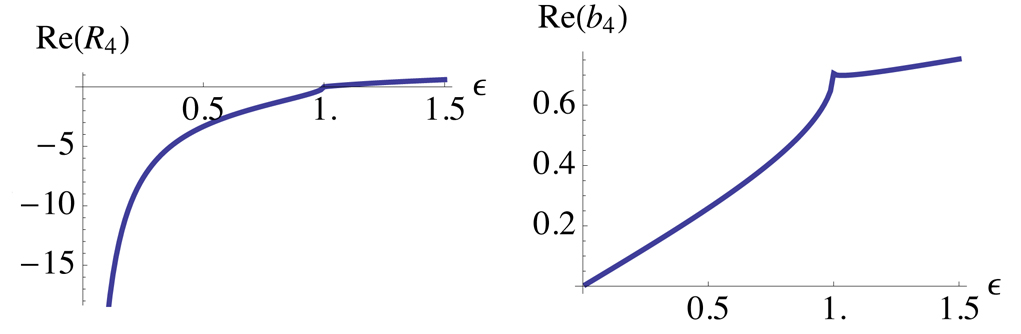}
\end{center}
\caption{Plots of the real parts of $R_4$ and $b_4$ for $0\leq\epsilon\leq1.5$.
For this plot $\gamma=0.05$ and $\omega=1.0$.}
\label{F14}
\end{figure}

Finally, we consider $a_4$ in (\ref{E33}). In Fig.~\ref{F13} we plot $a_4$ as a
function of $\epsilon$ and see that $a_4$ is real in the first broken-$\cPT$
region and in the unbroken-$\cPT$ region (\ref{E8}). Furthermore, $\Delta$ in
(\ref{E16}) is real and positive in the unbroken-$\cPT$ region. Thus, the
eigenvalues in (\ref{E32}) are real.  
In the unbroken region ${\rm Re}\,b$ in (\ref{E14}) is positive and ${\rm Re}\,
R$ in (\ref{E35}) is negative (see Fig.~\ref{F14}). Thus, the eigenfunctions
vanish exponentially in pairs of $90^\circ$-Stokes wedges centered about the
real axis in the $u$ plane and centered about the imaginary axis in the $y$
plane.
Because $a_4$ is positive the eigenspectrum (\ref{E32}) is bounded below and
because the eigenfunctions vanish exponentially in the appropriate Stokes
wedges in the $u$ and $y$ planes, we again regard this case as physically
acceptable.

It is interesting but perhaps not surprising that the Hamiltonian (\ref{E4})
has two distinct physically allowed positive spectra, which correspond to the
choices $a=a_2$ and $a=a_4$. It is not surprising that in the context of $\cPT$
quantum mechanics one Hamiltonian can have two independent positive spectra.
This phenomenon was discussed previously for the case of the sextic quantum
mechanical Hamiltonian $H=p^2+x^6$ in Ref.~\cite{R26}. This sextic Hamiltonian
also has two positive spectra, which are associated with two distinct pairs of
Stokes wedges in which the eigenfunctions vanish exponentially.

\section{Concluding remarks}
\label{s5}

In this paper we have studied the behavior of a system of two coupled
oscillators, one with gain and the other with loss. If the gain and loss
parameters are equal, the system is $\cPT$ symmetric. Furthermore, it is
described by a Hamiltonian and thus the energy is exactly conserved. Both the
classical and the quantum systems exhibit two transitions at exactly the same
values of the coupling parameter $\epsilon=\epsilon_1$ and $\epsilon=
\epsilon_2$.

Specifically, if the coupling is smaller than a critical value $\epsilon_1$,
the system is not in equilibrium even though the energy is conserved. At the
classical level the lack of equilibrium manifests itself as complex frequencies
and exponentially growing and decaying modes; at the quantum level the lack of
equilibrium is associated with complex energy levels. Above the critical value
$\epsilon_1$ the system is in an unbroken-$\cPT$-symmetric phase; at the
classical level the system is in equilibrium and the oscillators exhibit Rabi
oscillations and at the quantum level the system exhibits not one, but two
independent sets of real spectra and associated eigenfunctions.

There is also a second transition point $\epsilon_2$ above which the system is
no longer in equilibrium and the quantum energy levels become complex. This
super-strong coupling regime is very hard to study at the classical experimental
level because it requires that the oscillators be strongly coupled, so strongly
coupled that they are likely to interfere with one another. However, at the
quantum level it might be possible to perform experiments using quantum optics
techniques and that can actually observe the second phase transition. Such
quantum experiments might also prove to be extremely interesting because they
may provide a platform on which to study quantum entanglement \cite{R23}.

In future work we will study systems of more than two coupled oscillators. The
phase structure of such systems is interesting; we have found that as the number
of coupled oscillators increases, there are more and more phase transition
points. 

\acknowledgments
MG is grateful for the hospitality of the Department of Physics at Washington
University. CMB thanks the U.S.~Department of Energy and MG thanks the INFN
(Lecce) for financial support.


\begin{thebibliography}{99}
\bibitem{R22} The minus sign in (\ref{E5}) is optional; it is included so that
this definition of parity reduces to the conventional one for the case of just
one oscillator.
\end{thebibliography}
\end{document}